\documentclass[12pt]{article}
\usepackage[english]{babel}
\usepackage{amsmath,amsthm}
\usepackage{amsfonts}
\usepackage{latexsym}
\usepackage{graphicx}
\usepackage{graphics}
\usepackage{float}
\usepackage{hyperref}
\usepackage{amssymb}
\usepackage{authblk}
\title{Fundamental Physics and General Relativity with the LARES and LAGEOS satellites}

\author[1,2]{Ignazio Ciufolini\thanks{ignazio.ciufolini@unile.it}}
\author[3]{Antonio Paolozzi}
\author[4]{Rolf Koenig}
\author[5]{Erricos C. Pavlis}
\author[6]{John Ries}
\author[7]{Richard Matzner}
\author[8]{Vahe Gurzadyan}
\author[9]{Roger Penrose}
\author[3]{Giampiero Sindoni}
\author[3]{Claudio Paris}
\affil[1]{Dip. Ingegneria dell'Innovazione, Universit\`a del Salento, Lecce, Italy}
\affil[2]{Centro Fermi, Rome, Italy}
\affil[3]{Scuola di Ingegneria Aerospaziale, Sapienza Universit\`a di Roma, Italy}
\affil[4]{Helmholtz Centre Potsdam GFZ German Research Centre for Geosciences, Potsdam, Germany}
\affil[5]{Goddard Earth Science and Technology Center (GEST), University of Maryland, Baltimore County, USA}
\affil[6]{Center for Space Research, University of Texas at Austin, USA}
\affil[7]{Center for Relativity, University of Texas at Austin, USA}

\begin{document}
\maketitle

\begin{abstract}
Current observations of the universe have strengthened the interest to further test General Relativity and other theories of fundamental physics. After an introduction to the phenomenon of frame-dragging predicted by Einstein's theory of General Relativity, with fundamental astrophysical applications to rotating black holes, we describe the past measurements of frame-dragging obtained by the LAGEOS satellites and by the dedicated Gravity Probe B space mission. We also discuss a test of String Theories of Chern-Simons type that has been carried out using the results of the LAGEOS satellites. We then describe the LARES space experiment. LARES was successfully launched in February 2012 to improve the accuracy of the tests of frame-dragging, it can also improve the test of String Theories. We present the results of the first few months of observations of LARES, its orbital analyses show that it has the best agreement of any other satellite with the test-particle motion predicted by General Relativity. We finally briefly report the accurate studies and the extensive simulations of the LARES space experiment, confirming an accuracy of a few percent in the forthcoming measurement of frame-dragging.
\end{abstract}
\section{Introduction}

	The current study of the universe and nature has on the one hand allowed us to understand some of the basic laws governing the infinitely small, down to spatial distances of the order of $10^{-16}$ cm, or less, corresponding to quarks, and on the other hand to observe and in part understand the evolution of the universe up to spatial distances of the order of $10^{28}$ cm corresponding to the position of some quasars and near the so-called big-bang. With regard to the time scale, the current study of the universe ranges from a few instants after the big-bang up to the present time, approximately 14 billion years later. The laws of physics cover the four fundamental interactions: gravitational, electromagnetic, weak and strong. The last three are encompassed in the Standard Model theory of gauge symmetries.
	Theories, not yet experimentally verified, such as String and Brane-World theories, try to unify gravitation with the other three interactions and to unify the two great physical theories of General Relativity and Quantum Mechanics. The goal is the unification of the four interactions of nature in a theory that can be experimentally tested, and that can also address one of the biggest mysteries and riddles of science, the composition of most of the universe in which we live, that is the nature of dark energy and dark matter. Indeed, the discovery of the accelerated universe \cite{riess98,perl99} is one of the outstanding events in science today and dark energy, or ``quintessence'', is regarded as a new exotic physical substance that is accelerating the expansion of the universe. Dark energy, together with dark matter, should constitute approximately 95\% of the mass-energy of the universe in an unexplained form \cite{perl03,cald04,plan13}.

	The evolution of the universe and the gravitational interaction are currently described by Einstein's gravitational theory of General Relativity \cite{mtw}. General Relativity is a triumph of classical thought, created by Einstein to satisfy the competing requirements of the Equivalence Principle (local inertial physics can show no evidence of gravity), and the large scale effects of gravity. Einstein's gravitational theory succeeded by postulating that gravitation is the curvature of spacetime and it is a fundamental component for understanding the universe that we observe. During the past century General Relativity achieved an experimental triumph \cite{tury09,will,ciuw95}. On the one hand, a number of key predictions of Einstein's gravitational theory have been experimentally confirmed with impressive accuracy. On the other hand, General Relativity today has practical applications in space research, geodesy, astronomy and navigation in the Solar System, from the Global Navigation Satellite Systems (GNSS) to the techniques of Very Long Baseline Interferometry (VLBI) and Satellite Laser Ranging (SLR), and is a basic ingredient for understanding astrophysical and cosmological observations such as the expanding universe and the dynamics of binary systems of neutron stars.

	Despite being a well verified description of gravity, General Relativity has encountered somewhat unexpected developments in observational cosmology and is affected by some theoretical problems. Indeed, the study of distant supernovae in 1998 led to a discovery that they accelerate away from us. Since then, what is now referred to as dark energy is at the center of attention of many theoreticians. Observational data currently support its interpretation as the cosmological constant introduced by Einstein. However its current value, comparable with the critical density, needs to be reconciled with the expectations of quantum field theory or analogous fundamental theory (e.g., \cite{kam}). Combining gravity with quantum field theory may be expected to reveal the nature of dark energy and hence resolve the mystery of its value, and whether it might be related to dark matter.

	Among its theoretical problems, General Relativity predicts the occurrence of spacetime singularities \cite{pen65}, events in which every known physical theory ceases to be valid, the spacetime curvature diverges and time ends. Furthermore, General Relativity is a classical theory that does not include Quantum Mechanics and no one has succeeded in a quantized version of General Relativity, though this is a serious ongoing effort, with both Loop Quantum Gravity and String Theory approaches.
	Even though a breakdown of General Relativity should occur at the quantum level, some viable modifications of Einstein's theory already give different predictions at the classical level and might explain the riddle of the dark energy. Modifications of General Relativity on cosmological scales, for instance the so called f(R) theories (with higher order curvature terms in the action), have been proposed to explain the acceleration of the universe without dark energy \cite{def10}.
	In summary, every aspect of Einstein's gravitational theory should be directly tested and the accuracy of the present measurements of General Relativity and of the foundations of gravitational theories should be further improved.

\section{Frame-dragging}
\label{sec:1}

The observational tests of gravitational physics divide into purely solar system measurements of various effects, binary pulsars observations, and intermediate- and long- range cosmological observations via gravitational radiation. Purely solar system measurements include redshift and clock measurements, light deflection, time-delay of electromagnetic waves, Lunar Laser Ranging (LLR), geodetic precession and frame-dragging measurements.

Today, among the main challenges in experimental gravitation, we have the direct detection of gravitational waves, the improved measurement of the Post-Newtonian parameters testing General Relativity versus alternative gravitational theories, improved tests of the Equivalence Principle and the accurate measurement of frame-dragging and gravitomagnetism. Frame-dragging, or dragging of inertial frames, and gravitomagnetism and are produced by mass-energy currents, e.g., by the angular momentum of a body, in the same way as magnetism is generated by electric-currents in electrodynamics \cite{ciu07}.

The origin of inertia has intrigued scientists and philosophers for centuries and the inertial frames are at the foundations of physics and General relativity. What determines an inertial frame? In the Newtonian gravitational theory an inertial frame has an absolute existence, uninfluenced by the matter in the Universe. In Einstein's gravitational theory the {\it local} inertial frames have a key role \cite{mtw,wei,ciuw95}. The strong equivalence principle, at the foundations of General Relativity, states that the gravitational field is locally 'unobservable' in the freely falling frames and thus, in these local inertial frames, all the laws of physics are the laws of Special Relativity. However, the local inertial frames are determined, influenced and dragged by the distribution and flow of mass-energy in the Universe. The axes of these local, inertial frames are determined by free-falling torque-free test-gyroscopes, i.e., sufficiently small and accurate spinning tops. Therefore, these gyroscopes are dragged by the motion and rotation of nearby matter \cite{mtw,wei,ciuw95}, i.e., their orientation changes with respect to the distant stars: this is the `dragging of inertial frames' or `frame-dragging', as Einstein named it in a letter to Ernst Mach \cite{ein}. Frame-dragging represents in Einstein's theory the remnant of the ideas of Mach on the origin of inertia. Mach thought that centrifugal and inertial forces are due to rotations and accelerations with respect to all the masses in the Universe and this is known as Mach's principle \cite{ciuw95}.

In General Relativity, a torque-free spinning gyroscope defines an axis non-rotating relative to the local inertial frames, however, the orbital plane of a test particle is also a kind of gyroscope. Frame-dragging also has an intriguing influence on the flow of time and on electromagnetic waves propagating around a spinning body. Indeed, synchronization of clocks all around a closed path near a spinning body is not possible \cite{lanl,zeln} in any rigid frame not rotating relative to the `fixed stars', because light corotating around a spinning body would take less time to return to a starting point (fixed relative to the `distant stars') than would light rotating in the opposite direction \cite{lanl,zeln,ciur1,ciur2,ciurk}. Since frame-dragging affects clocks, light, gyroscopes \cite{pug,sch} (e.g., the gyroscopes of GP-B space experiment) and orbiting particles \cite{len} (see sections 4 and 5 on the LAGEOS satellites and on the LARES space experiment), it also affects matter orbiting and falling on a spinning body. Indeed, an explanation of the constant orientation of the spectacular jets from active galactic nuclei and quasars, emitted in the same direction during a time that may reach millions of years, is based on frame-dragging of the accretion disk due to a super-massive spinning black hole \cite{bard,tho} acting as a gyroscope.

The precession ${\dot {\bf \Omega}}_{Spin}$ of the spin axis of a test-gyroscope by the angular momentum $\bf J$ of the central body is: ${\dot {\bf \Omega}}_{Spin} = {{3 G (({\bf J} \cdot \hat{r}) \hat{r} - {\bf J})} \over {c^2 {r^3}}}$, where $\hat{r}$ is the position unit-vector of the test-gyroscope and r is its radial distance from the central body.
Similarly to a small gyroscope, the orbital plane of a planet, moon or satellite is a huge gyroscope that feels general relativistic effects. Indeed, frame-dragging produces a change of the orbital angular momentum vector of a test-particle, i.e., the Lense-Thirring effect, that is, the precession of the nodes of a satellite, i.e., the rate of change of its nodal longitude: ${\dot {\bf \Omega}}_{Lense-Thirring} = {{2 G \bf {J}} \over {c^2 a^3 (1-e^2)^{3/2}}}$, where $\bf \Omega$ is the longitude of the nodal line of the satellite (the intersection of the satellite orbital plane with the equatorial plane of the central body), $\bf J$ is the angular momentum of the central body, $a$ the semi-major axis of the orbiting test-particle, $e$ its orbital eccentricity, $G$ the gravitational constant and $c$ the speed of light. A similar formula also holds for the rate of change of the longitude of the pericentre of a test--particle, that is, of the so-called Runge-Lenz vector \cite{len,ciuw95}.

Frame-dragging phenomena, which are due to mass currents and mass rotation, may be usefully described by a formal analogy of General Relativity, in a weak gravitational field and for slow motion, with electrodynamics  (see Fig. 1) \cite{tho,ciuw95} and have been called gravitomagnetism. Whereas an electric charge generates an electric field and a current of electric charge generates a magnetic field, in Newtonian gravitational theory the mass of a body generates a gravitational field but a current of mass, for example the rotation of a body, would not generate any additional gravitational field. On the other hand, Einstein's gravitational theory predicts that a current of mass would generate a gravitomagnetic field that would exert a force on surrounding bodies and would change the spacetime structure by generating additional curvature \cite{ker}. Furthermore, in General Relativity a current of mass in a loop (that is, a gyroscope) has a behaviour formally similar to that of a magnetic dipole in electrodynamics, which is made of an electric current in a loop. Then, the gravitomagnetic field generates frame-dragging of a gyroscope, in a similar way to the magnetic field producing the change of the orientation of a magnetic needle (magnetic dipole). In General Relativity, the gravitomagnetic field, ${\bf {H}}$, due to the angular momentum ${\bf {J}}$ of a central body is, in the weak-field and slow-motion approximation: ${\bf {H}} ~ = ~ {\bf {\nabla}}~ \times ~ {\bf {h}} ~ \cong ~ 2 ~ G \Biggl [{{\bf J} \,- \,3 ({\bf J} ~ \cdot ~ {\hat x})~ {\hat {x}} \over {c^3 r^3}} \Biggr ]$, where $r$ is the radial distance from the central body, $\hat x$ is the position unit-vector and $\bf h$ is the so-called 'gravitomagnetic vector potential' (equal to the non-diagonal, space and time, part of the metric), see Fig. 1.

\begin{figure}[H]
\centering
\includegraphics[width=0.5\textwidth]{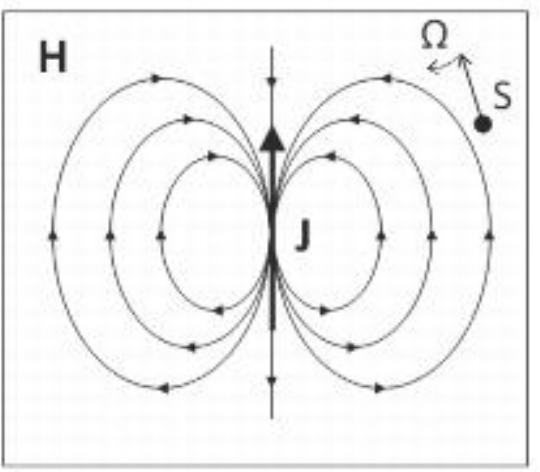}
\label{fig1}
 \caption{Frame-dragging and the gravitomagnetic analogy of General Relativity with electrodynamics. In General Relativity, freely falling test-gyroscopes define axes $fixed$ relative to the local inertial frames, where the equivalence principle holds, that is, where the gravitational field is locally `unobservable'; if we would rotate with respect to these gyroscope, we would then feel centrifugal forces, even though we may not rotate at all with respect to the `distant stars', contrary to our everyday intuition. Indeed, a gyroscope is dragged by spinning masses, that is, its orientation changes with respect to the `distant stars'. In this figure we show the gravitomagnetic field \cite{sch} $\bf H$ generated by the spin $\bf J$ of a central body and frame dragging $\bf {\dot \Omega}$ of a test gyroscope $\bf S$.}
\end{figure}

Since frame-dragging is due to the additional spacetime curvature produced by the rotation of a mass, to precisely characterize these phenomena, it has been proposed to use spacetime curvature invariants built using the Riemann curvature tensor (see: \cite{ciu1,ciu10na} and section 6.11 of \cite{ciuw95}). For discussions on the meaning of frame-dragging and gravitomagnetism, see: \cite{ashb,ocon,ocon2,ocon3,mnt,kop,mnt2,ciu1,ciu10na} and section 6.11 of \cite{ciuw95}.

\section{String Theories and the LAGEOS and LARES Satellites}

Among the extensions of General Relativity, the Chern-Simons gravity \cite{gur1} with the Pontryagin density coupled scalar field in the Einstein-Hilbert action, has attracted particular attention, since  Chern-Simons gravitational term also emerges from String theories and Loop Quantum Gravity (see e.g. \cite{gur2,gur3} and references therein). The Pontryagin scalar is:$ ^{\ast} R ^{\alpha \beta \mu \nu }  \, R_{\alpha \beta \mu \nu}$ , that is a pseudoinvariant built ``multiplying'' the Riemann tensor $R ^{\alpha \beta\mu \nu }$ with its dual $ ^{\ast} R ^{\alpha \beta \mu \nu } \equiv {1 \over 2} \varepsilon ^{\alpha \beta \sigma \rho} R _{\sigma \rho} ~ ^{\mu \nu}$, where $\varepsilon ^{\alpha \beta \sigma \rho}$ is the Levi Civita pseudotensor \cite{Petrov}. Due to the general character of Chern-Simons terms, the coupling constants can be even informative about the electroweak and even Planck scales \cite{gur3}, therefore any experimental constraint and even a null one can be of particular interest.

Concerning applications, Chern-Simons gravity has been involved to the interpretation of such basic cosmological and astrophysical problems as the dark energy, inflation, the evolution of binary neutron stars, gravitational wave emission by binary back holes and even the accretion powered energetic activity in the galactic nuclei and quasars  \cite{gur3,gur4,gur5,gur6,gur7}.
Astrophysical observations, however, still do not allow to obtain constraints on Chern-Simons terms, e.g., as it is in the case even for the binary pulsar J0737-3039 [4], and  the frame-dragging measurements near Earth are currently the only reasonable means to constraint the theory.

In 2008, Smith et al. \cite{gur8} showed that String Theories of the type of Chern-Simons gravity predict an additional drift of the nodes of a satellite orbiting a spinning body and of a gyroscope spin axis. Then, using the frame-dragging measurement obtained with the LAGEOS satellites, they constrained the coupling constant of Chern-Simons theory (which may also be related to dark energy and quintessence, and to more fundamental parameters, such as related to a quintessence field).  In particular, they set the lower limit to the Chern-Simons mass: $\vert m_{CS} \vert \, {\buildrel > \over \sim} \,  0.001 km^{-1}$.  See Fig. 2. Higher accuracy measurements by the LARES satellite will enable to improve that limit \cite{gur8,gurc13}.

\begin{figure}[H]
\centering
\includegraphics[width=0.5\textwidth]{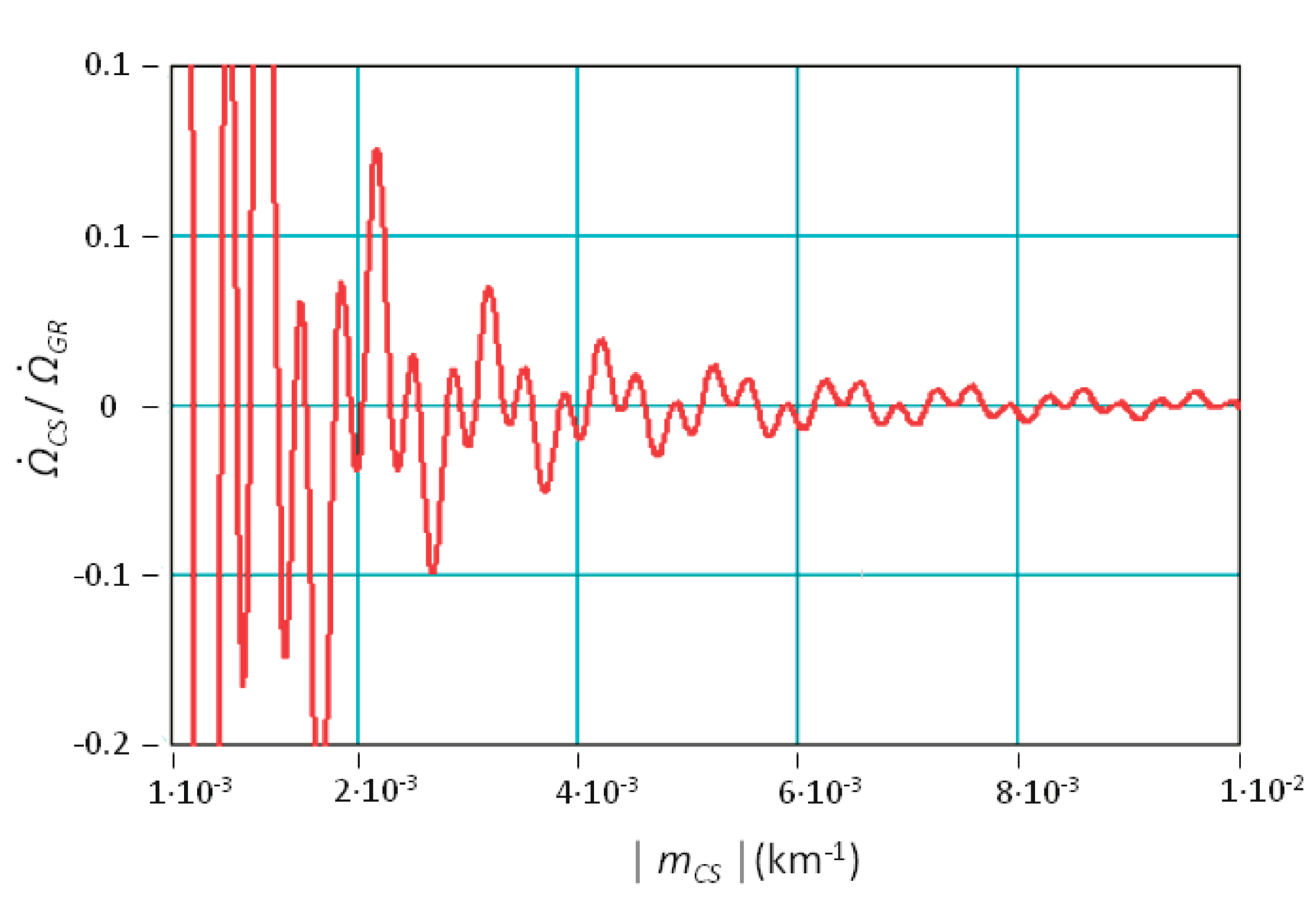}
\label{fig2}
 \caption{The ratio of the nodal rate of the LAGEOS satellites predicted by Chern-Simons gravity over that predicted by General Relativity implying the lower limit on the Chern-Simons mass: $\vert m_{CS} \vert \, {\buildrel > \over \sim} \,  0.001 km^{-1}$ (adapted from \cite{gur8}).}
\end{figure}

\section{Tests of Frame-Dragging with the LAGEOS satellites and Gravity Probe-B}

Since 1896 researchers, influenced by the ideas of Ernst Mach, tried to measure the frame-dragging effects generated by the rotation of the Earth on torsion balances \cite{frie} and gyroscopes \cite{fop}. In 1916, on the basis of General Relativity, de Sitter derived the Mercury perihelion precession due to the Sun angular momentum and, in 1918, Lense and Thirring \cite{len} gave a general weak-field description of the frame-dragging effect on the orbit of a test-particle around a spinning body, today known as Lense-Thirring effect (see section 4).
In 1959 and 1960, an experiment to test the general relativistic drag of a gyroscope was suggested \cite{pug,sch}. On 20 April 2004, after more than 40 years of preparation, the Gravity Probe B spacecraft was finally launched in a polar orbit at an altitude of about 642 km. The Gravity Probe B mission \cite{GPB} (see http://einstein.stanford.edu/) consisted of an Earth satellite carrying four gyroscopes and one telescope, pointing at the guide star IM Pegasi (HR8703), and was designed to measure the drifts predicted by General Relativity (frame-dragging and geodetic precession) of the four test-gyroscopes with respect to the distant `fixed' stars. General Relativity predicts that the average frame-dragging precession of the four Gravity Probe B’s gyroscopes by the Earth’s spin is about 39 milliarcseconds per year (that is 0.000011 degrees per year) about an axis contained in Gravity Probe B's polar orbital plane.
On 14 April 2007, after about 18 months of data analysis, the first Gravity Probe B results were presented: the Gravity Probe B experiment was affected by large drifts of the gyroscopes' spin axes produced by classical torques on the gyroscopes. The Gravity Probe B team explained \cite{buch} (see also \cite{baroco}) the large drifts of the gyroscopes as being due to electrostatic patches on the surface of rotors and housings, and estimated the unmodeled systematic errors to be of the order of 100 milliarcseconds per year, corresponding to an uncertainty of more than 250\% of the frame-dragging effect by the Earth spin.
In 2011, finally, the Gravity Probe B team claimed that by some modeling of the systematic errors they were able to reduce the uncertainty in the measurement of frame-dragging to 19 \% \cite{GPB}.
Frame-dragging is extremely small for Solar System objects, so to measure its effect on the orbit of a satellite we need to measure the position of the satellite to extremely high accuracy. Laser-ranging is the most accurate technique for measuring distances to the Moon and to artificial satellites such as LAGEOS (LAser GEOdynamics Satellite) \cite{lag}. Ultrashort-duration laser pulses are emitted from lasers on Earth and then reflected back to the emitting laser-ranging stations by retro-reflectors on the Moon or on artificial satellites. By measuring the total round-trip travel time of a laser pulse we are today able to determine the instantaneous distance of a retro-reflector on the LAGEOS satellites with a precision of a few millimeters \cite{noon} and their nodal longitude with an uncertainty of a fraction of a milliarcsec per year \cite{nasasi,ries,pet}.
In 1976, LAGEOS was launched by NASA and, in 1992, LAGEOS 2 was launched by the Italian Space Agency and NASA. They have altitudes of approximately 5,900 km and 5,800 km respectively. The LAGEOS satellites' orbits can be predicted, over a 15-day period, with an uncertainty of just a few centimeters \cite{nasasi,ries,pet}. The Lense-Thirring drag of the orbital planes of LAGEOS and LAGEOS 2 is \cite{ciu86,ciu89} approximately 31 milliarcseconds per year, corresponding at the LAGEOS altitude to approximately 1.9 m per year. Since using laser-ranging we can determine their orbits with an accuracy of a few centimeters, the Lense-Thirring effect can be measured very accurately on the LAGEOS satellites' orbits if all their orbital perturbations can be modeled well enough \cite{ciu86,ciu89,nasasi}. On the other hand, the LAGEOS satellites are very heavy spherical satellites with small cross-sectional areas, so atmospheric particles and photons can only slightly perturb their orbits and especially they can hardly change the orientation of their orbital planes \cite{ciu89,nasasi,rub,luc}.
By far the main perturbation of their orbital planes is due to the Earth's deviations from spherical symmetry and by far the main error in the measurement of frame-dragging using their orbits is due to the uncertainties in the Earth's even zonal spherical harmonics \cite{kau}. The Earth's gravitational field and its gravitational potential can be expanded in spherical harmonics and the even zonal harmonics are those harmonics of even degree and zero order. These spherical harmonics, denoted as $J_{2n}$, where ${2n}$ is their degree, are those deviations from spherical symmetry of the Earth's gravitational potential that are axially symmetric and that are also symmetric with respect to the Earth's equatorial plane: they produce large secular drifts of the nodes of the LAGEOS satellites. In particular, the flattening of the Earth's gravitational potential, corresponding to the second degree zonal harmonic $J_2$ describing the Earth's quadrupole moment, is by far the largest error source in the measurement of frame-dragging since it produces the largest secular perturbation of the node of LAGEOS \cite{ciu86,ciu96}. But thanks to the observations of the geodetic satellites, the Earth's shape and its gravitational field are extremely well known. For example, the flattening of the Earth's gravitational potential is today measured \cite{rolf} with an uncertainty of only about one part in $10^7$ that is, however, still not enough to test frame-dragging.
To eliminate the orbital uncertainties due to the errors in the Earth's gravity models, the use of both LAGEOS and LAGEOS2 was proposed \cite{ciu96}. However, it was not easy to confidently assess the accuracy of some earlier measurements \cite{sci} of the Lense-Thirring effect with the LAGEOS satellites, given the limiting factor of the uncertainty of the gravity models available in 1998.
In March 2002, the problem of the uncertainties in the Earth's gravity field was overcome when the twin GRACE (Gravity Recovery And Climate Experiment) \cite{gra1,gra2} spacecraft of NASA were launched in a polar orbit at an altitude of approximately 400 km and about 200-250 km apart. The spacecraft range to each other using radar and they are tracked by the Global Positioning System (GPS) satellites. The GRACE satellites have greatly improved our knowledge of the Earth's gravitational field. Indeed, by using the two LAGEOS satellites and the GRACE Earth gravity models, the orbital uncertainties due to the modeling errors in the non-spherical Earth's gravitational field are only a few per cent of the Lense-Thirring effect \cite{ciu04,ciu10,ciu11}. The method to measure the Lense-Thirring effect is to use two {\it observables}, provided by the two nodes of the two LAGEOS satellites, for the two unknowns: Lense-Thirring effect and uncertainty in the Earth quadrupole moment $\delta J_2$ \cite{ciu96}.
 In 2004, nearly eleven years of laser-ranging data were analyzed. This analysis resulted in a measurement of the Lense-Thirring effect with an accuracy \cite{ciu04,ciu07,ciu10,ciu11} of approximately 10\%. The uncertainty in the largest Earth's even zonal harmonic, that is the quadrupole moment $J_2$, was eliminated by the use of the two LAGEOS satellites, see Fig. 3. However, the main remaining error source was due to the uncertainty in the Earth even zonal harmonics of degree strictly higher than two and especially to the even zonal harmonic of degree four, i.e., $J_4$. After 2004, other accurate Earth gravity models have been published using longer GRACE observations. The LAGEOS analyses have then been independently repeated with new models, over a longer period and by using three different orbital programs developed by NASA Goddard, the University of Texas at Austin \cite{rie09}, see Fig. 4, and the German GeoForschungsZentrum (GFZ) Potsdam \cite{koe12}, see Fig. 5. The recent frame-dragging measurements \cite{ciu10,ciu11,koe12} by a team from the universities of Salento, Rome, Maryland, NASA Goddard, the University of Texas at Austin and the GFZ Potsdam, have confirmed the 2004 LAGEOS determination of the Lense-Thirring effect. No deviations from the predictions of General Relativity have been observed.

\begin{figure}[H]
\centering
\includegraphics[width=0.5\textwidth]{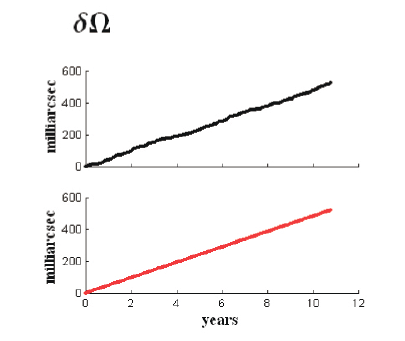}
\label{fig3}
\caption{The 2004 measurement of frame-dragging using the LAGEOS and LAGEOS 2 satellites \cite{ciu04,ciupavper}. The figure shows the {\it observed} orbital residuals of the nodal
longitudes, $\delta \Omega$, of the LAGEOS satellites, combined in a suitable way to eliminate the uncertainty of the Earth's quadrupole moment. In black is the raw, observed, residual nodal longitude of the LAGEOS satellites after
removal of six periodic signals. The best-fit line through these observed
residuals has a slope of 47.9 mas $yr^{-1}$. In red is the theoretical Lense-Thirring
prediction of Einstein's general relativity for the combination of the nodal
longitudes of the LAGEOS satellites; its slope is 48.2 milliarcsec $yr^{-1}$ (adapted from \cite{ciu04}).}
\label{fig:3}
\end{figure}

\begin{figure}[H]
\centering
\includegraphics[width=0.5\textwidth]{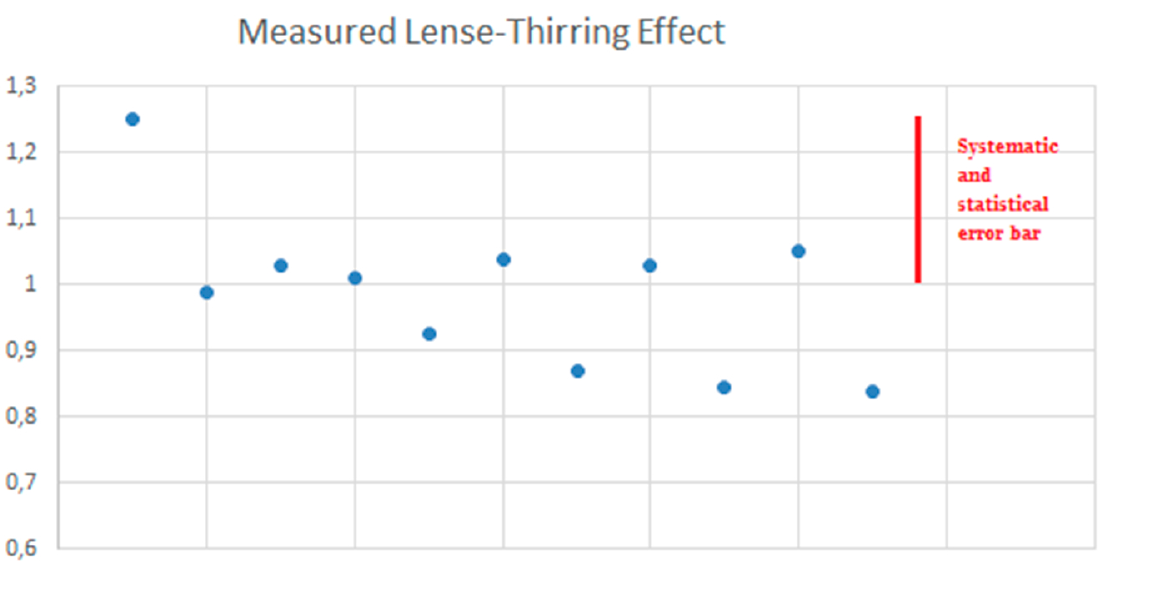}
\label{fig4}
\caption{Independent 2008 measurement of frame-dragging using LAGEOS and LAGEOS 2
obtaned by CSR of the University of Texas at Austin using UTOPIA and the GRACE
models: EIGEN-GRACE02S, GGM02S, EIGEN-CG03C, GIF22a, JEM04G,
EIGEN-GL04C, JEM01-RL03B, GGM03S, ITG-GRACE03S and EIGEN-GL05C. The mean value of frame-dragging measured by Ries et al. using these models is 0.99 of the prediction of General Relativity. The total error budget of CSR-UT in the measurement of frame-dragging is  about 12$\%$.; see \cite{rie09}.}
\label{fig:4}
\end{figure}

\begin{figure}[H]
\centering
\includegraphics[width=0.5\textwidth]{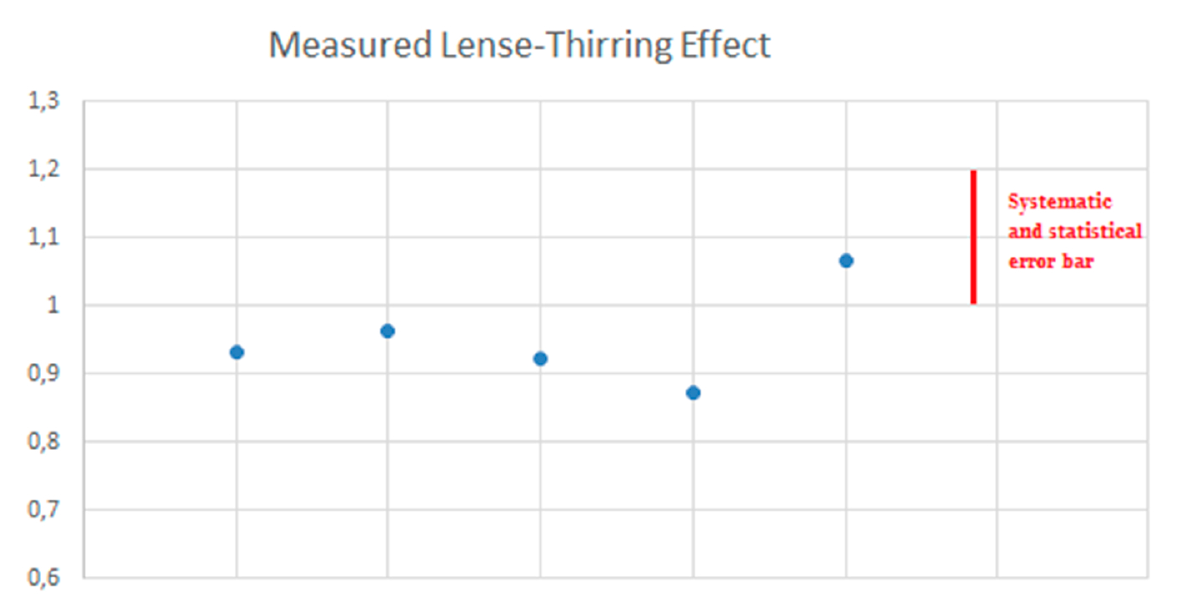}
\label{fig5}
 \caption{Independent 2012 measurement of frame-dragging using LAGEOS and LAGEOS 2
obtaned by GFZ Potsdam using EPOS-OC and the GRACE model: EIGEN-6C, EIGEN-6C (without considering trend and annual and semi-annual variations in the Earth gravitational field), EIGEN-6Sp.34, EIGEN-51C and EIGEN-GRACE03S. The mean value of frame-dragging measured by K\"onig et al. using these models is 0.95 of the prediction of General Relativity; see \cite{koe12}.}
\label{fig:5}
\end{figure}

\section{The LARES Space Experiment}

In the test of frame-dragging using LAGEOS and LAGEOS 2, the main error source is due to the even zonal harmonic of degree four, $J_4$; such an error can be as large as 10\% of the Lense-Thirring effect \cite{ciu10b}. Thus, to significantly increase the accuracy of the measurement of frame-dragging, one would need to eliminate that uncertainty by using an additional observable, i.e., by using a laser-ranged satellite in addition to LAGEOS and LAGEOS 2.

LARES (LAser RElativity Satellite) is a laser-ranged satellite of the Italian Space Agency (ASI), see Fig. 6. It was launched successfully on the 13th of February 2012 with the qualification flight of VEGA, the new launch vehicle of the European Space Agency (ESA), which was developed by ELV (Avio-ASI) \cite{ciu12,ciupao}. LARES, together with the LAGEOS and LAGEOS 2 satellites and the GRACE mission \cite{gra1,gra2}, will provide an accurate test of Earth's frame-dragging with uncertainty of a few percent and other tests of fundamental physics \cite{ciu10b,ciu11,ciu13}. The Lense-Thirring drag of the orbital planes of the LARES is approximately 118 milliarcseconds per year corresponding, at the LARES altitude, to approximately 4.5 m/yr.

The LARES orbital elements are as follows: the semi-major axis is 7820 km, orbital eccentricity 0.0007, and orbital inclination 69.5$^{o}$. It is currently successfully tracked by the global International Laser Ranging Service (ILRS) station network \cite{ILRS}. LARES has the highest mean density of any known object orbiting in the Solar System. It is spherical and covered with 92 retro-reflectors, and it has a radius of 18.2 cm. It is made of a tungsten alloy, with a total mass of 386.8 kg, resulting in a ratio of cross-sectional area to mass that is about 2.6 times smaller than that of the two LAGEOS satellites \cite{ciupao}. Before LARES, the LAGEOS satellites had the smallest ratio of cross-sectional area to mass of any artificial satellite, such a ratio is critical to reduce the size of the non-gravitational perturbations. Indeed, the extremely small cross-sectional area to mass ratio of LARES, i.e. 0.00027 $m^2/kg$, and its special structure, a single piece solid sphere with high thermal conductivity, ensure that the unmodeled non-gravitational orbital perturbations are smaller than for any other satellite, in spite of its lower altitude compared to LAGEOS. This behavior has been confirmed experimentally using the first few months of laser ranging observations \cite{ciu12}.

\begin{figure}[H]
\centering
\includegraphics[width=0.5\textwidth]{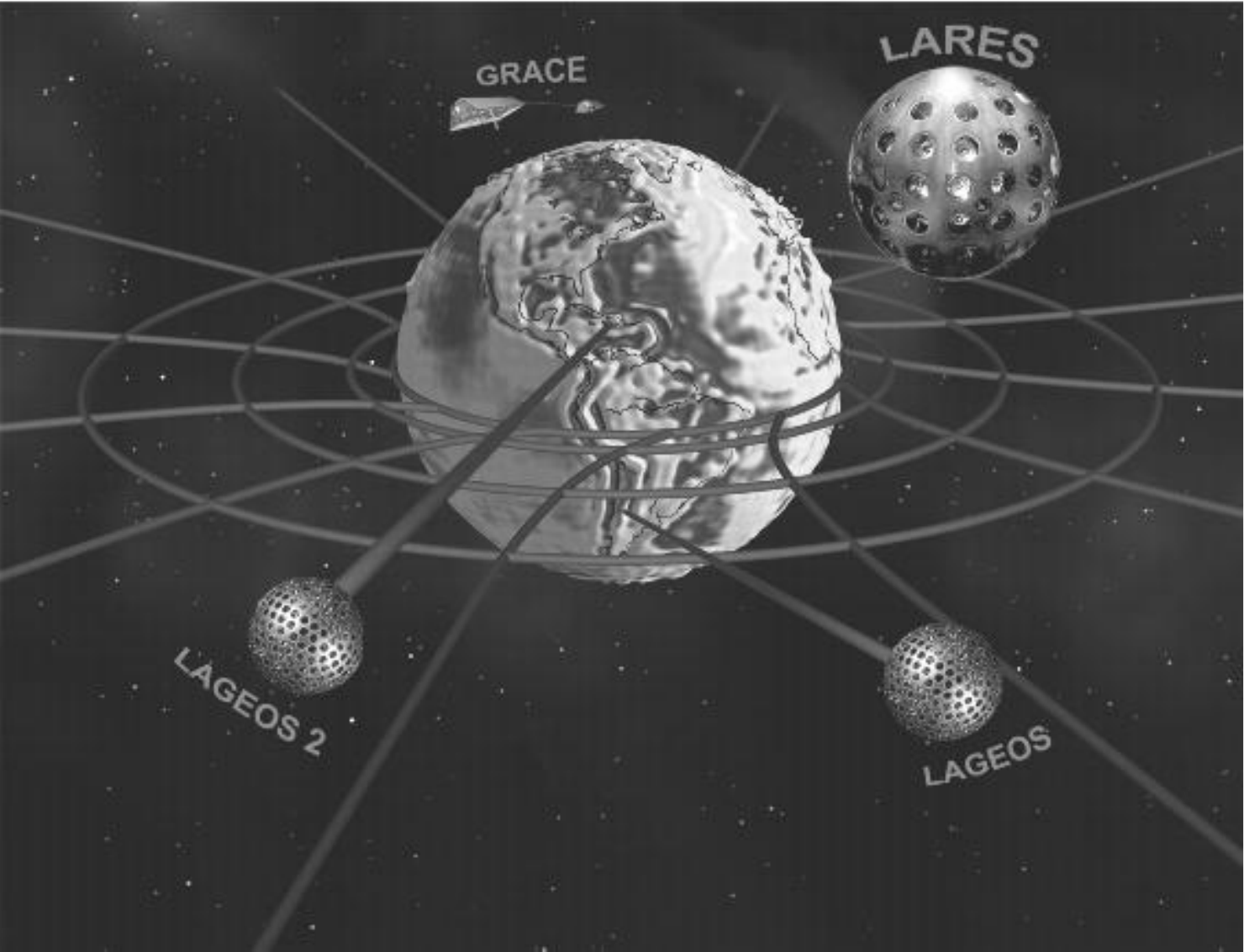}
\label{fig6}\caption{Artistic view of the LARES space experiment with the satellites: LARES, LAGEOS, LAGEOS 2 and GRACE. The radial twisted curves are an artistic representation of the spacetime twist owed to frame-dragging by the Earth rotation. The Earth is displayed using the gravitational field determination EIGEN-GRACE02S obtained with GRACE.}
\end{figure}

\subsection{First results of LARES orbital analysis}

At the very foundation of General Relativity is the geodesic motion of a small, structureless test-particle. Depending on the physical context, a star, planet or satellite can behave very nearly like a test-particle, so geodesic motion is used to calculate the advance of the perihelion of a planet's orbit, the dynamics of a binary pulsar system and of an Earth-orbiting satellite (a timelike geodesic path in spacetime's Lorentzian geometry is one that locally maximizes proper time, in analogy with the length-minimizing property of Euclidean straight lines). Verifying geodesic motion is then a test of paramount importance to General Relativity and other theories of fundamental physics.

General Relativity explains the gravitational interaction as the curvature of spacetime generated by mass-energy and mass-energy currents via the Einstein field equations \cite{mtw,haw75,ciuw95}. For example, the gravitational attraction of Earth on its Moon and artificial satellites is explained by General Relativity via the spacetime curvature generated by the Earth's mass. The motion of any test body within the gravitational field of another massive body, e.g., the motion of a `small' satellite around the Earth, is simply determined by a geodesic of spacetime with curvature generated by the massive body. Moon and artificial Earth satellites follow approximately geodesics of the spacetime with deviations from an ideal geodesic path due their finite size and to the non-gravitational forces acting on them. Thus, geodesic motion is at the foundation of General Relativity and of any other theory where the gravitational interaction is described by spacetime curvature dynamically generated by mass-energy. Therefore, the creation of the best possible approximation for the free motion of a test-particle, a spacetime geodesic, is a profound goal for experiments dedicated to the study of the spacetime geometry in the vicinity of a body, yielding high-precision tests of General Relativity and constraints on alternative gravitational theories.

A fundamental issue regards the approximation to a geodesic that is provided by the motion of an actually extended body. In General Relativity \cite{har03,rin01} the problem of an extended body is subtle, due not only to the non-linearity of the equations of motion, but also to the need to deal with the internal structure of the compact body, constructed of continuous media, where kinetic variables and thermodynamic potentials are involved. Further, there may be intrinsically non-local effects arising from the internal structure of the extended body, such as tidal influences. Moreover, there are problems concerning the approximations that need to be made in order to describe a given extended body as a test-particle moving along a geodesic. These problems are related to the fact that many of the common Newtonian gravitational concepts such as the `center of mass', `total mass' or `size' of an extended material body do not have well-defined counterparts in General Relativity \cite{ehl73}. The Ehlers-Geroch theorem \cite{ehl04} (generalizing the result in \cite{ger75}) attributes a geodesic to the trajectory of an extended body with a small enough own gravitational field, if for a Lorentzian metric the Einstein tensor satisfies the so-called dominant energy condition \cite{haw75}, this tensor being non-zero in some neighborhood of the geodesics and vanishing at its boundaries. This theorem, asserting that small massive bodies move on near-geodesics, thus achieves a rigorous bridge from General Relativity to space experiments with `small' satellites which suggests a high level of suppression of non-gravitational and self-gravitational effects from the satellite's own small gravitational field. This enables us to consider the satellite's motion to be nearly geodesic and hence provides a genuine testing ground for General Relativity's effects.

Given the extreme weakness of the gravitational interaction with respect to the other interactions of nature, the space environment is the ideal laboratory to test gravitational and fundamental physics. However, in order to test gravitational physics, a satellite must behave as nearly as possible as a test-particle and must be as little as possible affected by non-gravitational perturbations such as radiation pressure and atmospheric drag. In addition, its position must be determined with extreme accuracy.

The best realization of an orbiting test-particle is LARES. By measuring the total round-trip travel time of a laser pulse, it is possible to determine the instantaneous distance to the satellite with an accuracy of a few millimeters. However, in order to test gravitational physics, we not only need to measure the position of a body with extreme accuracy, but we also need it to behave like a test-particle. In space, a test-particle can be realized in two ways: a small drag-free satellite or a small spacecraft with high density and an extremely small area-to-mass ratio. In the case of the drag-free Gravity Probe-B satellite, a mean residual acceleration of about $40 \times 10^{-12} \, m/{s^2}$ was achieved [21]. For a passive satellite (with no drag-free system), the key characteristic that determines the level of attenuation of the non-gravitational perturbations is the density, reflected by the ratio between its cross-sectional area and its mass.

We processed the LARES laser ranging data based on the first seven 15-day arcs using the orbital analysis and data reduction systems UTOPIA of UT/CSR (Center for Space Research of The University of Texas at Austin), GEODYN II of NASA Goddard, and EPOS-OC of GFZ (Helmholtz Centre Potsdam GFZ German Research Centre for Geosciences) \cite{EPOSOC}. In all cases, state-of-the art satellite orbital dynamical models were employed, including all the general relativistic post-Newtonian corrections, GRACE-based mean gravity field models \cite{gra1,gra2}, modern models for the ocean and solid Earth tides, as well as solar radiation pressure, Earth albedo and atmospheric drag \cite{pav,mar,rub}. No `thermal thrust' \cite{rub1,rub2} models were used. For the 105 days analyzed, GEODYN, UTOPIA and EPOS-OC independently determined that the residual along-track accelerations for LARES were only about $0.4 \times 10^{-12} \, m/{s^2}$, whereas for the two LAGEOS satellites, the acceleration residuals were 1-$2 \times 10^{-12} \, m/{s^2}$.

\begin{figure}[H]
\centering
\includegraphics[width=0.5\textwidth]{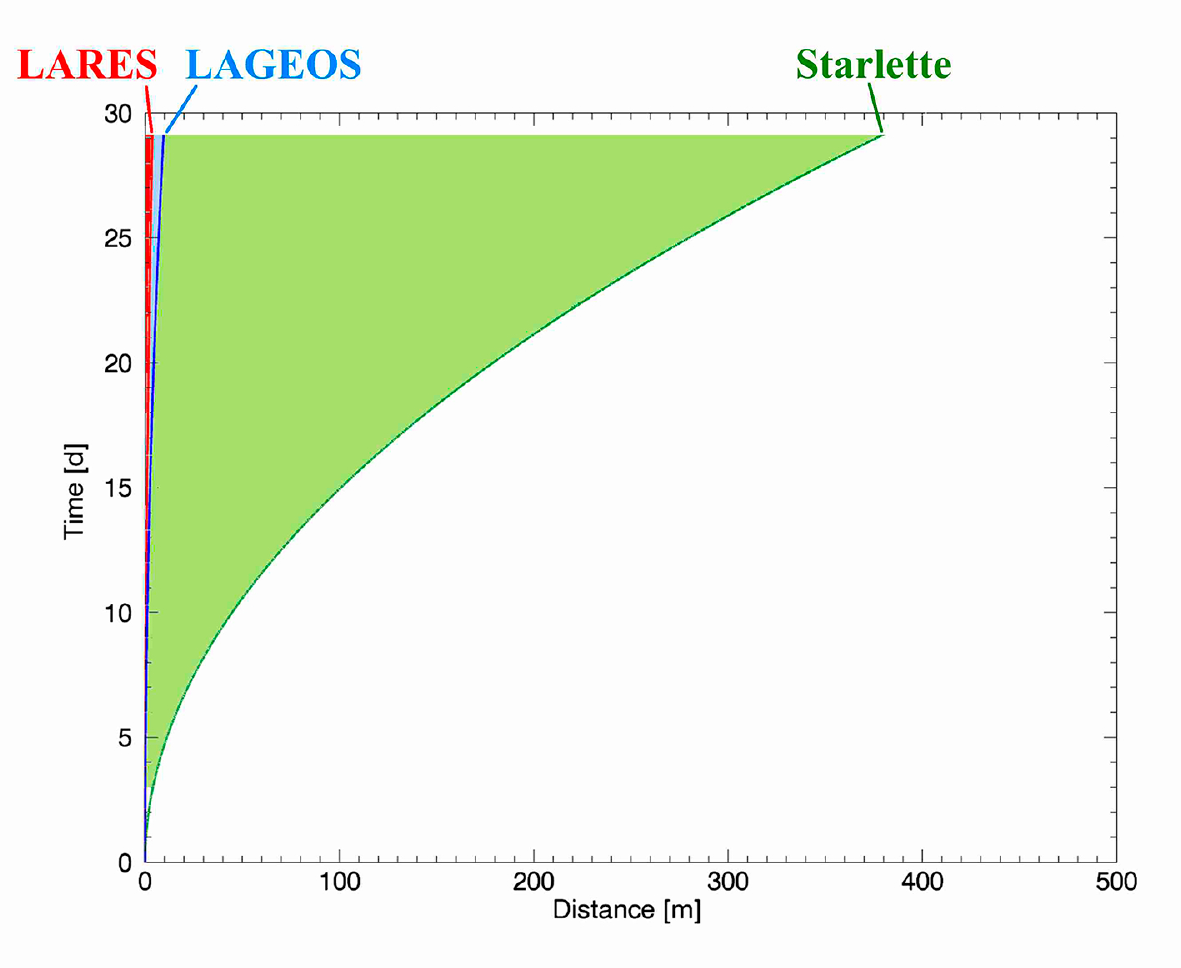}
\label{fig7} \caption{The red curve represents the change of distance between a `test-particle' following a spacetime geodesic, represented here by the axis of ordinates in a frame co-moving with the test-particle, and a similar particle perturbed by the average unmodelled along-track acceleration of the magnitude observed on the LARES satellite of approximately $0.4 \times 10^{-12} \, m/{s^2}$. The blue and green curves represents the change of distance between a test-particle and a similar particle perturbed by an average along-track acceleration of the typical size of the unmodelled along-track acceleration observed on the LAGEOS satellites, of the order of $1 \times 10^{-12} \, m/{s^2}$, and respectively of STARLETTE, with a typical residual acceleration of the order of $40 \times 10^{-12} \, m/{s^2}$. The axis of ordinates may be thought of to represent a spacetime geodesic followed by LARES or LAGEOS after removing all the known and unmodelled non-gravitational perturbations (adapted from \cite{ciu12}).}
\end{figure}

This is particularly impressive given that LARES is far lower in the Earth's atmosphere than LAGEOS. The residual along-track accelerations of a satellite provide a measure of the level of suppression of its non-gravitational perturbations: atmospheric drag, solar and terrestrial radiation pressure and thermal-thrust effects. Atmospheric drag acts primarily along the satellite's velocity vector, while solar radiation pressure, terrestrial radiation pressure (the visible and infrared radiation from Earth) and thermal-thrust effects will all have some contribution along-track as well. We recall that the Yarkovsky effect on a spinning satellite is a thermal thrust resulting from the anisotropic temperature distribution over the satellite's surface caused by solar heating. A variation of this effect, due to the Earth's infrared radiation, is the Earth-Yarkovsky or Yarkovsky-Rubincam effect \cite{rub1,rub2}.

The effects of the residual unmodelled along-track acceleration on the orbits of the laser ranged satellites: LARES, LAGEOS and STARLETTE (a CNES laser ranged satellite launched in 1975) are illustrated in fig. 7 where we plot the change in the distance from their `ideal' orbit, caused by the unmodelled along-track accelerations \cite{ciu12}. The vertical axis may be thought of as representing an `ideal' reference world line of LARES, LAGEOS and Starlette, `ideal' in the sense that all of its orbital perturbations are known. Figure 7 shows the unmodelled deviations from geodesic motion for LARES, LAGEOS and Starlette (once the known non-gravitational perturbations are removed, to the extent permitted by our current models) due to the unmodelled along-track accelerations. In these figures, we show the effect of a typical residual unmodelled along-track acceleration of $1 \times 10^{-12} \, m/{s^2}$ for LAGEOS, $0.4 \times 10^{-12} \, m/{s^2}$ for LARES and $40 \times 10^{-12} \, m/{s^2}$ for Starlette. Since all the general relativistic post-Newtonian corrections were included in our orbital analyses, these figures show the level of agreement of the LARES and LAGEOS orbits with the geodesic motion predicted by General Relativity.

It must be stressed that a residual unmodelled out-of-plane acceleration, constant in direction, of the order of magnitude of the unmodelled along-track acceleration observed on LARES, will produce an extremely small secular variation of the longitude of its node, i.e., of its orbital angular momentum. For example, by considering an out-of-plane acceleration with amplitude of $0.4 \times 10^{-12} \, m/{s^2}$ constant in direction, its effect on the node of LARES would be many orders of magnitude smaller than the tiny secular drift of the node of LARES due to frame-dragging [30] of about 118 milliarcsec/y. Therefore, LARES, together with the LAGEOS satellites, and with the determination of Earth's gravitational field obtained by the GRACE mission, will be used to accurately measure the frame-dragging effect predicted by General Relativity, improving by about an order of magnitude the accuracy of previous frame-dragging measurements by the LAGEOS satellites \cite{ciu04,ciu10,ciu11}.

In conclusion, LARES provides the best available test-particle in the Solar System for tests of gravitational physics and General Relativity, e.g., for the accurate measurement of frame-dragging, and after modelling its known non-gravitational perturbations, its orbit shows the best agreement of any satellite with the geodesic motion predicted by General Relativity.

\subsection{Error analysis and Monte Carlo Simulations of the LARES experiment}

A large number of papers have been published that analyze all the error sources, of both gravitational and non-gravitational origin, that can affect the LAGEOS and LARES experiments (see, e.g., \cite{ciu89,nasasi,ries,ciu96,pet,ciupavper,ciu10,ciu10b,ciu11,ciu13,gurc13}. The largest measurement uncertainties are due to the errors in the first two Earth even zonal harmonics, of degree 2 and 4, i.e., $\delta J_2$ and $\delta J_4$, but they are eliminated using three observables, i.e., the three nodes of the LARES, LAGEOS and LAGEOS 2 satellites, thus allowing a measurement of frame-dragging with an uncertainty of a few percent. Furthermore, the LARES inclination of 69.5$^{o}$ minimizes the uncertainties due to the error in the Earth even zonal harmonics of degree higher than four, i.e., $\delta J_{2n}$ with $2n > 4$. This is the largest source of error in the measurement of frame-dragging using the LAGEOS. LAGEOS 2 and LARES satellites. The error in the LARES experiment due to each even zonal harmonic up to degree 70 was analyzed in detail in \cite{ciu10b,ciu11}. The LARES error analyses have been recently confirmed by a number of Monte Carlo simulations \cite{ciu13}.

In Fig. 8 we display the error in the LARES experiment due to each even zonal harmonic up to degree 70. In this figure, the largest errors due to the uncertainties in the first two even zonal harmonics, of degree 2 and 4, are not shown since they are eliminated in the measurement of frame-dragging using the 3 observables, i.e. the 3 nodes of LARES, LAGEOS and LAGEOS 2. Fig. 8 clearly displays that the error due to each even zonal harmonic of degree higher than 4 is considerably less than 1\% and in particular that the error is substantially negligible for the even zonal harmonics of degree higher than 26.

The results of Fig. 8 are based on the calibrated uncertainties (i.e., including systematic errors) of the EIGEN-GRACE02S (GFZ, Potsdam, 2004) model (used in \cite{ciu04}). In Fig. 8 we also display the maximum percent errors due to each even zonal harmonic obtained by considering as uncertainty for each harmonic the difference between the value of that harmonic in the EIGEN-GRACE02S model minus its value in the GGM02S model (a model with comparable accuracy); this is a standard technique in space geodesy to estimate the reliability of the published uncertainties of a model; of course, in order to use this technique, one must use models of comparable accuracy, i.e., models that are indeed comparable, or use this technique only to assess the errors of the less accurate model.

Using EIGEN-GRACE02S and GGM02S (see \cite{ciu10b}), the total error in the measurement of the Lense-Thirring effect due to the even zonal harmonics is respectively 1.4 \% and 2.1 \%. Even though the real error in the EIGEN-GRACE02S coefficients would probably be about two or three times larger than these published uncertainties, EIGEN-GRACE02S was just a preliminary 2004 determination of the Earth gravitational field and models much more accurate than EIGEN-GRACE02S, based on much longer GRACE observations. are today available. Indeed, these two models, EIGEN-GRACE02S and GGM02S have been obtained with a relatively small amount of observations of the GRACE spacecraft (launched in February 2002) and therefore a substantial factor of improvement over these two GRACE models has to be taken into account at the time of the LARES data analysis (between 2012 and 2018), thanks to longer GRACE observational periods and to other space geodesy missions too.

\begin{figure}[H]
\centering
\includegraphics[width=0.5\textwidth]{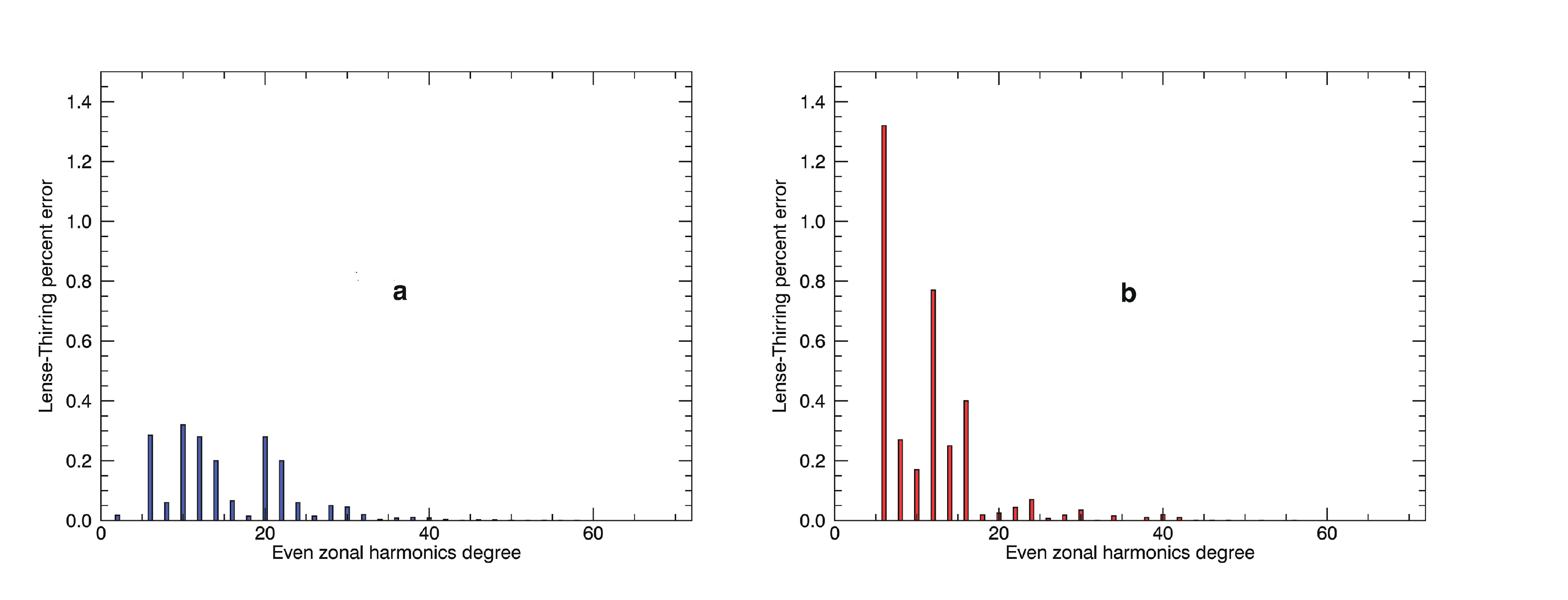}
\label{fig8}
\caption{Percent error in the measurement of frame-dragging using LARES, LAGEOS and LAGEOS 2 as a function of the uncertainty due to each even zonal harmonic. The points in blue in panel $a$ are the errors obtained using the model EIGEN-GRACE02S and the points in red in panel $b$ are the errors obtained using as uncertainty of each coefficient the difference between the value of this coefficient in the two different models EIGEN-GRACE02S and GGM02S. The total error in the measurement of the Lense-Thirring effect using EIGEN-GRACE02S is 1.4 \% and by using as uncertainties the differences between the coefficients of the two models is 3.4 \%.  However, at the time of the LARES data analysis a substantial improvement has to be taken into account with respect with these older 2004 models that were based on less than 365 days of observations of the GRACE  spacecraft. Today, the GRACE determinations of the Earth gravitational field are already much more accurate than the two 2004 GRACE models used to derive the Earth gravitational field displayed in figure 8.}

\end{figure}

In regard to a detailed treatment of the other orbital perturbations that affect the LARES experiment, tidal effects and non-gravitational perturbations, such as solar and albedo radiation pressure, thermal thrust and particle drag, we refer to \cite{ciu89,nasasi,ciupavper,ciu10,ciu10b}. In regard to the orbital perturbations on the LARES experiment due to the time dependent Earth's gravity field, we observe that the largest tidal signals are due to the zonal tides with $l=2$ and $m=0$, due to the Moon node, and to the $K_1$ tide with $l=2$ and $m=1$ (tesseral tide). However, the error due to the medium and long period zonal tides ($l=2$ and $m=0$) will be eliminated, together with the static $J_{2}$ error, using the combination of the three nodes (also the uncertainties in the time-dependent secular variations $\dot J_{2}$, $\dot J_{4}$ will be cancelled using this combination of three observables). Furthermore, the tesseral tide $K_1$ will be fitted for over a period equal to the LARES nodal period (see \cite{nasasi} and chapter 5 of \cite{pet}) and this tide would then introduce a small uncertainty in our combination.
In regard to the non-gravitational orbital perturbations, we simply observe here that the LAGEOS satellites and especially the LARES satellite are extremely dense spherical satellites with very small cross-sectional-to-mass ratio in order to reduce their non-gravitational perturbations \cite{ciu89}. In particular, in the previous section 5.1 we have shown that the unmodelled perturbations of the LARES orbit, in spite of its lower orbit, are smaller than on the LAGEOS satellites owed to the much smaller cross-sectional-to-mass ratio of LARES and to its special structure. We finally point out that the neutral and charged particle drag on the LARES node is a negligible effect. That is owed to the almost circular orbit of LARES, i.e., its orbital eccentricity is $e \cong 0.0007$, and to the LARES special structure. Indeed, even assuming that the exosphere would be co-rotating with the Earth at any satellite altitude, in the case of zero orbital eccentricity, $e=0$, the total nodal shift of the satellite would be zero, as calculated in \cite{ciu89}. Indeed, the nodal rate of a satellite due to particle drag is a function of $sin \; \nu \cdot cos  \; \nu$ (where $\nu$ is the true anomaly) and the total nodal shift is then zero over one orbit. In the case of a very small orbital eccentricity, the total nodal shift would be proportional to the eccentricity and thus for LARES it would be a very small effect \cite{ciu89} owed also to its very small cross-sectional-to-mass ratio.

A number of Monte Carlo simulations have recently confirmed the previous detailed and extensive error analyses of the LARES experiment \cite{ciu13}, i.e., the potentiality of the LARES experiment to achieve a measurement of frame-dragging with an uncertainty of a few percent only. These simulations have confirmed that the three observables provided by the three nodes of the LARES, LAGEOS and LAGEOS 2 satellites, together with the latest Earth gravitational field determinations from the GRACE space mission, will allow us to improve significantly the previous measurements of the phenomenon of frame-dragging predicted by General Relativity, by eliminating the uncertainties in the value of the first two even zonal harmonics of the Earth potential $\delta J_2$ and $\delta J_4$.

The 100 simulations were designed to reproduce as closely as possible the real experiment to measure frame-dragging using LARES, LAGEOS, LAGEOS-2 and GRACE. We considered a number of physical parameters whose uncertainties have a critical impact on the accuracy of the measurement of the frame-dragging effect using LARES, LAGEOS and LAGEOS-2. Together with the values of these critical parameters, determined either by the GRACE space mission (in the case of the Earth gravitational field parameters) or by previous extensive orbital analyses (in the case of the radiation pressure parameters of the satellites), we consider their realistic uncertainty estimated by also taking into account the systematic errors. Then, using EPOS-OC, we simulated (100 times) the orbits of the LARES, LAGEOS and LAGEOS 2 satellites by randomly generating values of the GM (mass) of Earth, of its five largest even zonal harmonics, $J_2, \, J_4, \, J_6, \, J_8$ and $J_{10}$, of the secular rate of change of the two largest even zonal harmonics $\dot {J_2}$ and $\dot {J_4}$ and of the solar radiation coefficients of LARES, LAGEOS and LAGEOS 2. The frame-dragging effect was always kept equal to its General Relativity value. Finally, we carried out the analysis of their simulated laser-ranging observations.

The result of the 100 simulations of the LARES experiment was that the standard deviation of the measured simulated values of frame-dragging was equal to 1.4\% of the frame-dragging effect predicted by General Relativity. Its mean value effect was equal to 100.24\% of its general relativistic value. Thus, the Monte Carlo simulations confirmed an error budget of about 1\% in the forthcoming measurement of frame-dragging using LARES, LAGEOS, LAGEOS 2 and GRACE.

\section{Conclusions}

Frame-dragging is an intriguing phenomenon predicted by General Relativity with fundamental astrophysical applications to rotating black holes. Past measurements of frame-dragging have been performed using the LAGEOS satellites and the dedicated Gravity Probe B space mission, respectively with accuracies of about 10\% and 19\%. The LAGEOS tests of frame-dragging have been independently obtained by three teams: Universities of Salento, Sapienza and Maryland, University of Texas at Austin, and GFZ Potsdam, using three different orbital programs. The LAGEOS results were also used to constrain String Theories of Chern-Simons type. The LARES space experiment will improve the measurement of frame-dragging by one order of magnitude by also improving the test of String Theories. The orbital analyses of the first few months of observations of LARES have shown that the LARES orbit has the best agreement of any other satellite with the test-particle motion predicted by General Relativity. Accurate error analyses and extensive simulations have confirmed a total error of a few percent in the forthcoming measurement of frame-dragging using LARES, LAGEOS, LAGEOS 2 and GRACE.

\section{Acknowledgements}

The authors gratefully acknowledge the International Laser Ranging Service for providing high-quality laser ranging tracking of the LARES satellites. I. Ciufolini and A. Paolozzi gratefully acknowledge the support of the Italian Space Agency, grants I/043/08/0, I/016/07/0, I/043/08/1 and I/034/12/0. J.C. Ries the support of NASA Contract NNG06DA07C, and E.C. Pavlis and R.A. Matzner the support of NASA Grant NNX09AU86G.

\end{document}